\documentclass[aps,prl,groupedaddress,showpacs,floatfix]{revtex4-1}

\usepackage{graphicx}
\usepackage{comment}
\usepackage{amsmath}
\begin{document}
\def\be{\begin{equation}}
\def\ee{\end{equation}}
\def\ba{\begin{eqnarray}}
\def\ea{\end{eqnarray}}
\def\parderiv#1#2{\frac{\partial #1}{\partial #2}}

\def\rvec{{\bf r}}
\def\rvece{{\bf r_e}}
\def\rvecm{{\bf r_m}}
\def\omegavec{{\bf \omega}}
\def\Vvec{{\bf V}}
\def\Avec{{\bf A}}
\def\Rvec{{\bf R}}
\def\Dvec{{\bf D}}
\def\vvec{{\bf v}}
\title{A relativistic framework to establish coordinate time on the Moon and beyond}

\author{Neil Ashby\footnote{Neil.Ashby@Colorado.Edu}  \& Bijunath R. Patla\footnote{bijunath.patla@nist.gov}}
\affiliation{National Institute of Standards and Technology, Boulder, CO 80305}

\date{\today}

\begin{abstract}
As humanity aspires to explore the solar system and investigate distant worlds such as the Moon, Mars, and beyond, there is a growing need to establish and broaden coordinate time references that depend on the rate of standard clocks. 
According to Einstein's theory of relativity, the rate of a standard clock is influenced by the gravitational potential at the location of the clock and the relative motion of the clock. A coordinate time reference is established by a grid of synchronized clocks traceable to an ideal clock at a predetermined point in space. This allows for the comparison of local time variations of clocks due to gravitational and kinematic effects.
We present a relativistic framework to introduce a coordinate time for the Moon. This framework also establishes a relationship between the coordinate times for the Moon and the Earth as determined by standard clocks located on the Earth's geoid and the Moon's equator.
A clock near the Moon's equator ticks faster than one near the Earth's equator, accumulating an extra 56.02 microseconds per day over the duration of a lunar orbit.
This formalism is then used to compute the clock rates at  Earth-Moon Lagrange points. Accurate estimation of the rate differences of coordinate times across celestial bodies and their inter-comparisons using clocks onboard orbiters at relatively stable Lagrange points as time transfer links is crucial for establishing reliable communications infrastructure. This understanding also underpins precise navigation in cislunar space and on celestial bodies' surfaces, thus playing a pivotal role in ensuring the interoperability of various position, navigation, and timing (PNT) systems spanning from Earth to the Moon and to the farthest regions of the inner solar system.

\end{abstract}

\maketitle
















\section{1.~Introduction}
More than 50 years after the first lunar landing, a multinational consortium, which includes NASA, is working towards a return to the Moon under the Artemis Accords~\cite{artemis_20}. Our ability to explore distant worlds will require the design and development of a 
communication and navigation infrastructure within and beyond cislunar space. With the expectation of a significant increase in assets on the lunar surface and in cislunar space in the near future,  developing a robust architecture for accurate position, navigation, and timing (PNT) applications has become a matter of paramount interest.

Communication and navigation systems rely on a network of clocks that are synchronized to each other within a few tens of nanoseconds. 
As the number of assets on the lunar surface grows, synchronizing local clocks with higher precision using remote clocks on Earth becomes challenging and inefficient. An optimal solution would be to draw from the heritage of global navigation satellite systems (GNSS) by envisioning a system or constellation time common to all assets and then relating this time to clocks on Earth. 

The relativistic framework presented here enables us to compare clock rates on the Moon and cislunar Lagrange points with respect to clocks on Earth by using a metric appropriate for a locally freely falling frame. The time measured by a clock at any given location is known as the proper time. Relativity of simultaneity implies that no two observers will agree on a given sequence of events if they are in different reference frames~\cite{einstein96}. In other words, clocks in different reference frames tick at different rates. The gravitational and motional effects affect the ticking rate of clocks when compared with ``ideal" clocks that are at rest and sufficiently far away from any gravitating mass. For example, clocks farther away from Earth tick faster, and clocks in uniform motion will tick slower with respect to ``ideal" clocks, and vice-versa. Therefore, choosing an appropriate reference frame becomes essential for obtaining self-consistent results when comparing clocks on two celestial bodies. 

In this paper, mainly we seek answers to the following questions:  What is a good choice for the coordinate system that can be used to relate the proper times on the Earth and the Moon? What is an appropriate choice for the locations of ideal clocks on the surfaces of the Earth and Moon that makes it easier to compare their proper times? What is the proper time difference between clocks on the Moon and the Earth? What are the proper time differences between clocks located at the Earth-Moon Lagrange points and the Earth? 
The stability offered by Lagrange points provides a low acceleration noise environment for spacecraft with clocks.  The relativistic corrections for such clocks can be precisely estimated as their positions and velocities are well-determined and can be used to compare the proper times of clocks on Earth, Moon, and in cislunar orbits.

In Section~1, we use the global positioning system (GPS) as an example to illustrate the relativistic effects on clocks if the Moon is treated just like an artificial satellite of the Earth and obtain a rough estimate for the clock rates on the Moon with respect to clocks on the geoid. Section~2 introduces a freely falling coordinate system with its center coinciding with the center of mass of the Earth and Moon. Section~3 compares the rate offset of a clock on the lunar surface to clocks on the geoid using this freely falling coordinate system, assuming the Moon is in a Keplerian orbit around the Earth. The results are compared with precise orbits for the Moon obtained using the latest planetary ephemerides DE440~\cite{park_2021}.
Section~4 discusses the time rate offsets at Earth-Moon Lagrange points $L_1, L_2 $, and $L_4/L_5$. Conclusions and future outlook are presented in Section~5. Appendices 1 and 2 introduce the framework for developing the metric used in all calculations. Appendix 3 justifies our assumptions of using a Keplerian model ignoring tidal effects, and a discussion in Appendix~4 establishes general covariance, meaning that the results are coordinate-independent.


\section{2.~ Clocks in orbit}
An instructive example of establishing a coordinate time on Earth is the GPS time. The constellation clocks are set to beat at the average coordinate rate corresponding to clocks at rest on the surface of the rotating Earth by applying  a ``factory frequency offset" to the clocks before launch, which is~\cite{ashby03}
\be\label{offset}
\frac{\Delta f}{f}=\frac{3 GM_e}{2 a c^2}+\frac{\Phi_0}{c^2}\,,
\ee
where $a$ is the satellite semi-major axis, $G$ is Newton's gravitational constant, $M_e$ is the mass of the Earth, $c$ is the speed of light in vacuum, and $\Phi_0$ is the effective gravitational potential in the rotating frame, which is the sum of the static gravitational potential of the Earth, and a centripetal contribution~\cite{ashby03}. The International Astronomical Union defines a ``Terrestrial Time"(TT) by adopting a fixed value for $\Phi_0/c^2$~\cite{iau}: 
\be\label{LGdef}
-L_G=\frac{\Phi_0}{c^2}=-6.969290134(0) \times 10^{-10}\,.
\ee
If we simply substitute into Eq.~(\ref{offset}) the length of the semi-major axis of the Moon's orbit, $3.84748\times 10^8$ meters, the ``offset" becomes $-6.7964 \times 10^{-10}$.
To convert to a rate difference in microseconds per day, multiply by $86400 \times 10^6$ $\mu$s/day, yielding $58.721\ \mu{\rm s/day}$.
This does not include the effect of the Moon's gravitational potential. Also, this approach can be questioned because it does not make sense to treat the Moon's potential, from the point of view of an Earth-based inertial frame, as an Earth satellite;  the Moon's potential should be treated as a tidal potential.   Nevertheless, a standard clock on an Earth satellite at the distance of the Moon would beat faster than a standard clock at rest on Earth by 58.721~$\mu$s per day, not including any effect from the gravitational potential of the Moon.   This is a combination of Earth's gravitational potential and second-order Doppler shifts at the orbiting satellite. 

In addition, there are well-understood periodic effects arising from orbit eccentricity, with an  additional contribution to the rate on the satellite clock given by~\cite{ashby03}
\be 
\Delta=-\frac{2 GM_e}{c^2}\bigg(\frac{1}{a}-\frac{1}{r}\bigg)\,,
\ee
For a satellite in Keplerian orbit, the periodic contribution to the rate is
\be
\Delta = \frac{2GM_e e}{c^2 a (1-e^2)}(\cos(f) +e),
\label{gps_pert}
\ee
where $e$ is the eccentricity and $f$ is the true anomaly.
Numerical evaluation of Eq.~(\ref{gps_pert}) yields a value $1.2695 \times 10^{-12}(\cos(f)+e)$ or $0.1097 (\cos(f)+e)\ \mu {\rm s/day}$.
The time average of the combination $\cos(f) +e$ is zero, so this term does not contribute to the average rate.

This model is based on using an eccentric Keplerian orbit in the local inertial frame centered on Earth's center of mass.  
The center of mass of the Earth and Moon approximately follows a Keplerian orbit.  However, for the Earth-Moon system, one cannot have a Keplerian orbit in a coordinate system centered on the Earth and a Keplerian orbit in a coordinate system centered on the Earth-Moon center of mass with the same orbit parameters. There are also relativistic effects arising from changes in time and length scales, Lorentz contraction, and changes in tidal effects.

In the following sections, we shall investigate a local inertial system with an origin at the Earth-Moon center of mass. This is a freely falling inertial frame only in the Sun's gravitational field. 
The reason for using such a frame is that the Earth and Moon are treated more or less equivalently; tidal potentials are due only to the Sun. 
By addressing relativistic effects in this simple system, it may be expected that the main relativistic corrections can be better understood. 
We work in a plane containing the Earth, Moon, and Sun, which is inclined with respect to the ecliptic plane. 
Calculations are carried out only to order $c^{-2}$. Contributions from the tidal potentials of other solar system bodies are left out.  The metric signature is $(-1,1,1,1)$ with Greek indices running from 0 to 3, and a negative sign is assigned for gravitational potential rather than a positive sign used in geodesy.

\subsection{2.1~Coordinate Time}
In establishing a coordinate time on and near the Earth, two relativistic effects are compensated by adjusting the rates of standard clocks.  These are (a) the gravitational potential at the geoid and (b) the second-order Doppler shift due to the Earth's rotation.  
The gravitational potential at the equator can be estimated from existing models of Earth's gravitational potential.  
Viewed from an Earth-centered inertial frame, the second-order Doppler shift is
\be
\left(\frac{\delta f}{f}\right)_{\rm Dop}=-\frac{\omega_e^2 a_e^2}{2 c^2},
\ee
where $\omega_e$ is Earth's angular rotation rate and $a_e$ is the equatorial radius.  
%
Because the Earth's geoid is nearly a surface of effective hydrostatic equilibrium, all atomic clocks on the geoid beat at equal rates, and this rate can be calculated on the Earth's geoid at the equator.
The effective potential $\Phi_0$ in Eq.~(\ref{LGdef})
represents the fractional rate difference between an atomic clock at rest at infinity if the Earth were the only celestial body and an atomic clock fixed on the geoid of the rotating Earth.

A locally inertial, freely falling reference frame can be constructed at the center of mass of the Earth.   Such a construction (see Appendix 1 and 2) gives the following expression for the fundamental scalar invariant $ds^2$ to order $c^{-2}$~\cite{ashbybertotti86}:
\ba 
-ds^2=G_{\mu\nu}dx^{\mu}dx^{\nu}\hbox to 1 truein{}\nonumber\\
=-\left(1+\frac{2\Phi_e}{c^2}+\frac{2}{c^2}\big(\Phi_m+\Phi_s\big)\big|_{\rm tidal}\right)(dx^0)^2\nonumber \\
+\left(1-\frac{2\Phi_e}{c^2}-\frac{2}{c^2}\big(\Phi_m+\Phi_s\big)\big|_{\rm tidal}\right)
(dx^2+dy^2+dz^2)\,,
\ea 
where $dx^0, dx, dy$, and $dz$ are the changes in coordinate time and coordinate displacements. We have explicitly included the gravitational potential of the Earth, $\Phi_e$, and the tidal potentials of the Moon and Sun, $(\Phi_m+\Phi_s)|_{\rm tidal}$,  but have left out the small contributions from tidal potentials of other solar system bodies. If the time scale is adjusted by
\be 
dx^0 \rightarrow dx^0\left(1-\frac{\Phi_0}{c^2}\right)\,,
\label{scale_change}
\ee
then the scalar invariant becomes
\ba 
-ds^2=-\left(1+\frac{2(\Phi_e-\Phi_0)}{c^2}+\frac{2}{c^2}\big(\Phi_m+\Phi_s\big)\big|_{\rm tidal}\right)(dx^0)^2\nonumber\\
+\left(1-\frac{2\Phi_e}{c^2}-\frac{2}{c^2}\big(\Phi_m+\Phi_s\big)\big|_{\rm tidal}\right)
(dx^2+dy^2+dz^2)\,.
\ea 
With this adjustment in scale, apart from tidal effects which average to near zero, clocks at rest on the geoid beat at the rate of International Atomic Time (TAI), defined by atomic clocks at rest on the geoid.  Coordinate time suitable for use in navigation and timekeeping near the Earth's surface is then obtained by synchronizing clocks in the local inertial frame~\cite{ashby79}.
The proper time on a clock at rest on the geoid then, apart from tidal contributions, beats at the rate of coordinate time because the term $\Phi_0$ cancels the potential and second-order Doppler shifts on the geoid.


\subsection{2.2~Local Frame for the Moon}

While the Moon appears fairly rigid, it is nearly spherical due to hydrostatic equilibrium.  One can imagine a locally inertial, freely falling reference frame with its origin at the Moon's center of mass, see Appendix 1.  Near the Moon, the scalar invariant will be
\ba 
-ds^2=-\left(1+\frac{2\Phi_m}{c^2}+\frac{2}{c^2}\big(\Phi_e+\Phi_s\big)\big|_{\rm tidal}\right)(dx^0)^2\nonumber\\
+\left(1-\frac{2\Phi_m}{c^2}-\frac{2}{c^2}\big(\Phi_e+\Phi_s\big)\big|_{\rm tidal}\right)(dx^2+dy^2+dz^2)\,.
\ea 
Omitting tidal terms for the moment, a standard clock at rest at the Moon's equator will be subject to the gravitational potential of the Moon and to time dilation from the Moon's rotation.  Using a model of the Moon's potential~\cite{bert}  that includes spherical harmonics of degree and order up to 350, the gravitational potential on the Moon's equator and second-order Doppler shift, respectively, are found to be approximately $
\Phi_m\big|_{\theta=\pi/2}=-2.82101(7)\times 10^6~{\rm m}^2/{\rm s}^2,$ and $
-\omega_m^2 a_m^2/2=-10.70118(14)~{\rm m}^2/{\rm s}^2$,
where $a_m=1738140 (123)$~m is the equatorial radius and $\omega_m=2.661621\times 10^{-6}$/s is the sidereal rotation rate of the Moon. The Moon's rotation is tidally locked to the Earth.  Thus we could define a constant $L_m$ and a corresponding equipotential surface or ``selenoid" for the Moon:
\ba 
L_m =  -\frac{\Phi_{0m}}{c^2} = - \left( \frac{\Phi_m\big|_{\theta=\pi/2}}{c^2} - \frac{\omega_m^2 a_m^2}{2c^2} \right)\, 
= 3.13881(15)\times 10^{-11}\,.
\label{eq_lm}
\ea
We may then, in analogy to the time scale change for the Earth, define a new time scale for the Moon such that the scalar invariant near the Moon becomes
\ba 
-ds^2=-\left(1+\frac{2(\Phi_m-\Phi_{0m})}{c^2}+\frac{2}{c^2}\big(\Phi_e+\Phi_s\big)\big|_{\rm tidal}\right)(dx^0)^2\nonumber\\
+\left(1-\frac{2\Phi_m}{c^2}-\frac{2}{c^2}\big(\Phi_e+\Phi_s\big)\big|_{\rm tidal}\right)(dx^2+dy^2+dz^2)\,.
\ea 
Then, apart from tidal effects, standard clocks at rest on an effective equipotential of the rotating Moon will beat at equal rates and can be used to define the rate of coordinate time on the Moon:
$-ds^2 = -(dx^0)^2$.

\section{3.~Clock Rate Differences Between Earth and Moon}

The Earth and the Moon orbit around their mutual center of mass in different Keplerian orbits. Meanwhile, the center of mass of the Earth-Moon system orbits around the Sun in an approximately Keplerian orbit. To calculate the rate differences between clocks on Earth and on the Moon, a fictitious locally freely-falling inertial frame is introduced at the Earth-Moon center of mass. This makes it convenient to calculate the proper times elapsed on moving clocks in terms of Keplerian motions of the Earth and the Moon. The Sun's contribution is only tidal effects. If we omit the tidal potential of the Sun, the scalar invariant takes a simple form (Appendix 2)
\ba
-ds^2 = -\left(1+\frac{2\Phi_e}{c^2}+\frac{2 \Phi_m}{c^2}\right)(dx^0)^2
	+\left(1-\frac{2\Phi_e}{c^2}-\frac{2 \Phi_m}{c^2}\right)(dx^2+dy^2+dz^2)\,.
 \label{metric_cm}
\ea
Consider a clock fixed on the surface of the rotating geoid of Earth.  Since the geoid is a surface of approximate hydrostatic equilibrium, if such clocks are viewed from the local inertial frame they beat at the same rate, which can be evaluated at the equator.  The proper time on the Earth-based clock becomes
\ba\label{dtauEarth}
-c^2 d\tau_e^2=-\left(1+\frac{2\Phi_e}{c^2}\big|_{R_{Eeq}}+\frac{2\Phi_m}{c^2}\big|_{R_{Eeq}}\right)(dx^0)^2
+\frac{\big(\Vvec_e+\vvec_e\big)^2}{c^2} (dx^0)^2\,,
\ea
where the equatorial radius of the Earth is denoted by $\Rvec_{Eeq}$, $\Vvec_e$ is the velocity of the Earth's center of mass in the Earth-Moon coordinate system, and where $\vvec_e$ represents the velocity of the clock on the equator due to Earth rotation.  Expanding the velocity term, taking square roots of both sides and rearranging,
\ba 
c d\tau_e=\left(1+\frac{\Phi_e}{c^2}\big|_{R_{Eeq}}
-\frac{\vvec_e^2}{2 c^2}+\frac{\Phi_m}{c^2}\big|_{R_{Eeq}}
-\frac{V_e^2}{2 c^2}
-\frac{(\Vvec_e \cdot \vvec_e )^2}{c^2}\right)(dx^0)\,.
\ea
The first two contributions can be identified with the quantity $\Phi_0$.  The contribution from the Moon's potential can be well approximated by setting
\be 
\frac{\Phi_m}{c^2}\big|_{R_{Eeq}}=-\frac{GM_m}{c^2 D}\,,
\ee
where $D$ is the Earth-Moon distance.  The dot product term between velocities will depend on the specific position of the clock and will vary with a daily period; this variation is similar to the corrections to the gravitational potential contribution from the Moon arising from the fact that the clock is not at the center of the Earth.  Omitting such contributions gives
\be
c d\tau_e=\left(1+\frac{\Phi_0}{c^2}-\frac{GM_m}{c^2 D}-\frac{V_e^2}{2c^2}
\right)dx^0\,.
\ee
A similar argument applied to a clock fixed on the rotating Moon's surface of hydrostatic equilibrium gives the proper time 
\be 
c d\tau_m=\left(1+\frac{\Phi_{0m}}{c^2}-\frac{GM_e}{c^2 D}-\frac{V_m^2}{2c^2}
\right)dx^0\,,
\ee
where $\Vvec_m$ is the velocity of the Moon's center of mass, and $\Phi_m$, discussed above, is the combination of the Moon's gravitational potential on the selenoid, and second-order Doppler shift due to the rotation of a clock on the Moon's equator.  Therefore, the fractional frequency shift of a clock on the Moon's equator relative to a clock on Earth's equator is 
\ba
\frac{d\tau_m-d\tau_e}{d\tau_e}= \frac{(GM_m-GM_e)}{c^2 D}
+\frac{\Phi_{0m}-\Phi_0}{c^2}-\frac{1}{2 c^2}(V_m^2-V_e^2)\,,
\label{rate_Moon}
\ea
where $GM_e$ and $GM_m$ are the standard gravitational parameters for the Earth and Moon.
The distance to the Moon from the Earth, for a Keplerian orbit, is given by (see Appendix~3)
\be\label{DtoMoon}
D = \frac{a(1-e^2)}{1+e\cos(f)}\,,
\ee
where $f$ is the true anomaly plus possibly a constant, $a$ is the length of the semi-major axis, $M_T=M_e + M_m$, and  $e$ is the eccentricity of the Moon's orbit. 
Then 
\be 
\dot D = \frac{(1-e^2) a e \sin(f)}{(1+e \cos(f))^2}\frac{df}{dt}\,,
\ee
\be 
\dot f = \frac{df}{dt} = \frac{n(1+e\cos(f))^2}{(1-e^2)^{3/2}}\,,
\ee
\be 
n^2=\frac{GM_T}{a^3}\,.
\ee
The following combination of quantities occurs frequently and can be reduced to a simpler expression
\be\label{identity}
D^2 {\dot f}^2+(\dot D)^2=\frac{GM_T}{a}\left(\frac{1+2e \cos(f) +e^2}{1-e^2}\right)\,.
\ee
The velocities have radial as well as transverse components.  Consider first the quantity $\Vvec_e^2$.  The radial and transverse components of this velocity are 
\be 
V_r=-\frac{M_m}{M_T}\dot D\,, \quad V_t=\frac{M_m}{M_T} D \dot f.
\ee
Therefore
\ba
\frac{V_e^2}{2 c^2} = \frac{1}{c^2}\left(\frac{M_m^2}{ M_T^2}\right)(\dot D^2+D^2 \dot f^2)
 =  \mu^2 \frac{GM_T}{2 a c^2}\left(\frac{1+2e \cos(f) +e^2}{1-e^2}\right),
\ea
where $\mu=M_m/M_T=0.012150$.
A similar calculation for $V_m$ yields
\ba 
\frac{V_m^2}{2 c^2}=(1-\mu)^2\frac{GM_T}{2 a c^2}\left(\frac{1+2e \cos(f) +e^2}{1-e^2}\right)\,.
\ea
The difference of squares of velocities is then
\ba
\frac{V_e^2-V_m^2}{2 c^2}=(2\mu -1)\left(\frac{GM_T}{2 a c^2}\right)\left(\frac{1+2e \cos(f) +e^2}{1-e^2}\right)\,.
\label{rate_Moon_v2}
\ea
Using Eq.(\ref{rate_Moon_v2}) in Eq.(\ref{rate_Moon}), we obtain
\be
\frac{d\tau_m-d\tau_e}{d\tau_e}= \frac{(GM_m-GM_e)}{c^2 D}
+\frac{\Phi_{0m}-\Phi_0}{c^2}- (1-2\mu)\left(\frac{GM_T}{2 a c^2}\right)\left(\frac{1+2e \cos(f) +e^2}{1-e^2}\right).
\label{rate_Moon_f}
\ee
Now we'll discuss the small position-dependent terms that have been omitted. $\Dvec$ represents the vector from the center of the Earth to the center of the Moon.
The actual distance from the center of the Moon to a clock on Earth's geoid is $\vert -\Dvec+\rvec_e \vert$\,,
where $\rvec_e$ is the vector from the Earth's center to the clock on the equator.  Then
\ba
-\frac{GM_m}{c^2\vert -\Dvec+\rvec_e \vert}
 \approx -\frac{GM_m}{c^2 D}-\frac{GM_m}{c^2 D}\frac{\Dvec \cdot \rvece}{
D^2}\,.
\ea
The second term will vary with the rotation of the Earth.  Its magnitude is approximately
\be 
\frac{GM_m a_e}{c^2 D^2}\approx .0002\  \mu {\rm s/day}.
\ee
A similar term has been omitted from the calculation of the gravitational potential of the Earth on a Moon-fixed clock:
\be 
\frac{GM_e a_m}{c^2 D^2}\approx .0045\  \mu {\rm s/day}.
\ee
This term will have a period of approximately 27 days.  Cross terms between velocities of the center of mass and rotational velocities have been omitted.  One such term is
\be 
\frac{\Vvec_e \cdot \vvec_e}{c^2} \approx \frac{V_e \omega_e a_e}{c^2}\approx 0.0055 \  \mu {\rm s/day}.
\ee
This contribution will also vary with the rotation period of the Earth.  Another such term is
\be 
\frac{\Vvec_m \cdot \vvec_m}{c^2} \approx \frac{V_m \omega_m a_m}{c^2}\approx 0.0045 \  \mu {\rm s/day}.
\ee
Since the Moon is tidally locked to the Earth, its center of mass velocity and the velocity of a clock on the selenoid will be highly correlated.  Therefore, this term might give rise to a constant long-term average.  
\begin{table}
    \centering
    \begin{tabular}{ll}
    \hline
    \hline
       $GM_e$, Earth's gravitational parameter \hspace{4cm} & $3.986004418(8) \times 10^{14}\, {\rm m}^3{\rm s}^{-2} $~\cite{gmm} \\
       $GM_m$, Moon's gravitational parameter  & $4.90280031(44) \times 10^{12}\, {\rm m}^3{\rm s}^{-2} $~\cite{konopliv}\\
        $c$, speed of light in vacuum & $299792458(0)\, {\rm m}{\rm s}^{-1} $ \cite{const1,const2} \\
       $-\Phi_0/c^2, L_G $  & $ 6.969290134(0) \times 10^{-10}$, \quad $\sim 60.2~\mu${s/day}~\cite{iau} \\
       $-\Phi_{0m}/c^2, L_m $ & $ 3.13881(15)\times 10^{-11} $, \quad $\sim 2.71~\mu${s/day}, ~see Eq.~(\ref{eq_lm})\\
        $e$, assumed eccentricity of the Moon's orbit around the Earth & 0.05490~\cite{moon_fact} \\
        $a$, assumed Earth-Moon semi-major axis distance  & $ 3.84399 \times 10^8 $~m ~\cite{moon_fact}\\
        \hline
    \end{tabular}
    \caption{Constants and values.}
    \label{tab:const}
\end{table}
Omitting the position-dependent terms, we have used the constants listed in Table~\ref{tab:const} to evaluate the constant contribution and the amplitude of the periodic term. We find:
\be 
\frac{d\tau_m-d\tau_e}{d\tau_e}=6.48378(15)\times 10^{-10}-1.25502518(89) \times 10^{-12}\cos(f)\,.
\ee
Multiplying by $10^6 \times 86400$ $\mu$s to obtain a time difference per day gives
\be
56.0199(12) - 0.10843417(89)\cos(f)  \ \mu {\rm s/day}\,.
\ee

\begin{figure}
    \centering
    \includegraphics[width=\textwidth]{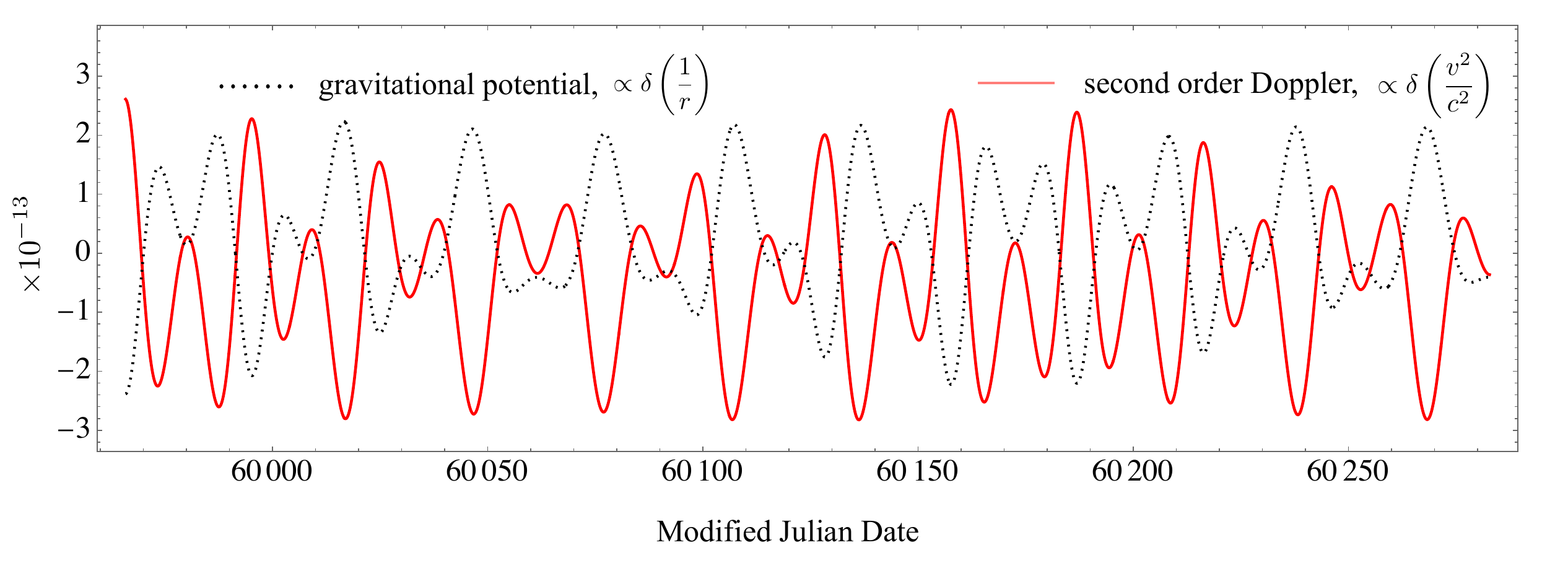}
    \caption{The residual of the gravitational potential and second-order Doppler for MJD~59965~(21 January 2023) to MJD~60282~(4 December 2023). $\Delta GM = GM_m - GM_e$ and $\Delta V^2 = V_m^2 - V_e^2 $, where the subscripts refer to the Moon and the Earth respectively, and the notation $\delta$ denotes the quantities are residuals with the Keplerian prediction subtracted from values computed using DE440. The deviation from an ideal Keplerian orbit due to tidal dynamics causes the Moon to move faster as the separation between the Earth and the Moon decreases. The uncertainty on the estimated rate of proper time on the Moon compared to the proper time of the Earth's geoid depends on how the above excursions are modeled and accounted for. Without clock corrections, the inaccuracy in time and position estimates can be as high as ten percent compared to clocks in well-determined orbits.}
    \label{fig:pot_dop_comp}
\end{figure}

\begin{figure}
    \centering
    \includegraphics[width=\textwidth]{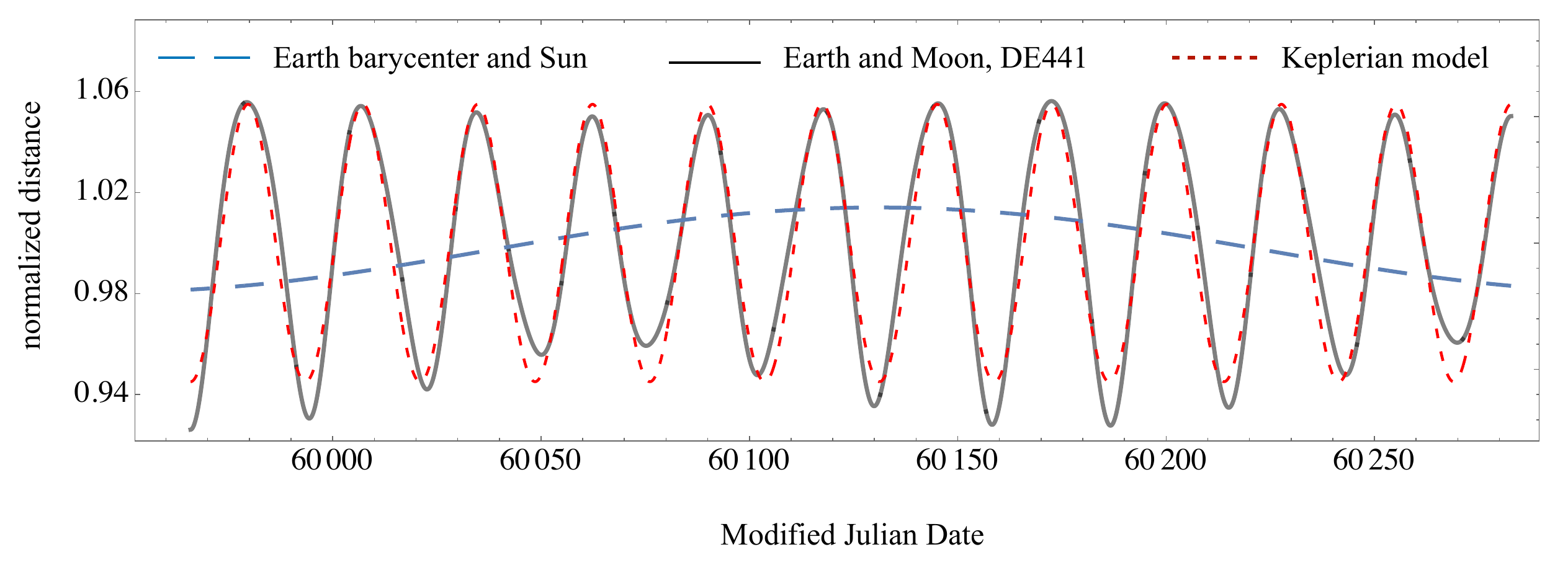}
    \caption{Earth-Moon distance compared with the Keplerian model.
    The normalized distance between the Earth and the Moon computed using the latest planetary ephemerides, DE440, is compared with the Keplerian model in the freely falling reference frame centered on the Earth-Moon barycenter. The tidal pull fluctuates at the perigee crossing for the Moon's orbit around the Earth as the Earth-Moon barycenter orbits the Sun. As a result, the actual Earth-Moon distance fluctuates compared to the Keplerian model at the perigee crossing. The tidal acceleration on the moon due to Earth and Sun is given in Eq.~(\ref{app1_tide}). DE440 accounts for the first term in  Eq.~(\ref{app1_tide}), whereas the Keplerian model doesn't include any tidal terms.  The second term in Eq.~(\ref{app1_tide}) is much smaller than the first term and is due to the Sun's effect on the Earth-Moon barycenter. 
    The phase offset between the Earth's orbit around the Sun and the amplitude modulation of the Earth-Moon distance is due to the inclination of the lunar orbit ($\sim 5^{\circ}$) with respect to the equatorial coordinate system, with the $xy$ plane coinciding with the Earth's equator.
    }
    \label{fig:distance_comp}
\end{figure}

None of the above estimates include tidal effects.  This omission is because as a tidal force pushes back and forth on a satellite, two other side effects have to be accounted for.  These are a change in the position of the satellite that entails a change in the gravitational potential of the body about which the satellite is orbiting and a change in the velocity of the satellite clock that changes its second-order Doppler shift. 
The residuals of the gravitational potential and second-order Doppler shift for the Earth-Moon system obtained by subtracting the Keplerian model from that obtained from DE440 are graphed in Fig.~\ref{fig:pot_dop_comp}.
Previous work on such problems has shown that these changes are of similar orders of magnitude. Summarizing, there are three contributions to the frequency shift of a clock in a satellite that are of similar orders of magnitude:
(1) the perturbing tidal potential itself;
(2) the perturbed position that changes the contribution from the main potential;
(3) the perturbed velocity that changes the time dilation contribution.
Although the perturbing tidal potential can easily be estimated, calculating the other two contributions is more complicated. When the Keplerian model is compared with DE440 ephemerides, the effects of solar tides are plotted in Fig.~\ref{fig:distance_comp}.



\begin{figure}
    \centering
    \includegraphics[width=\textwidth]{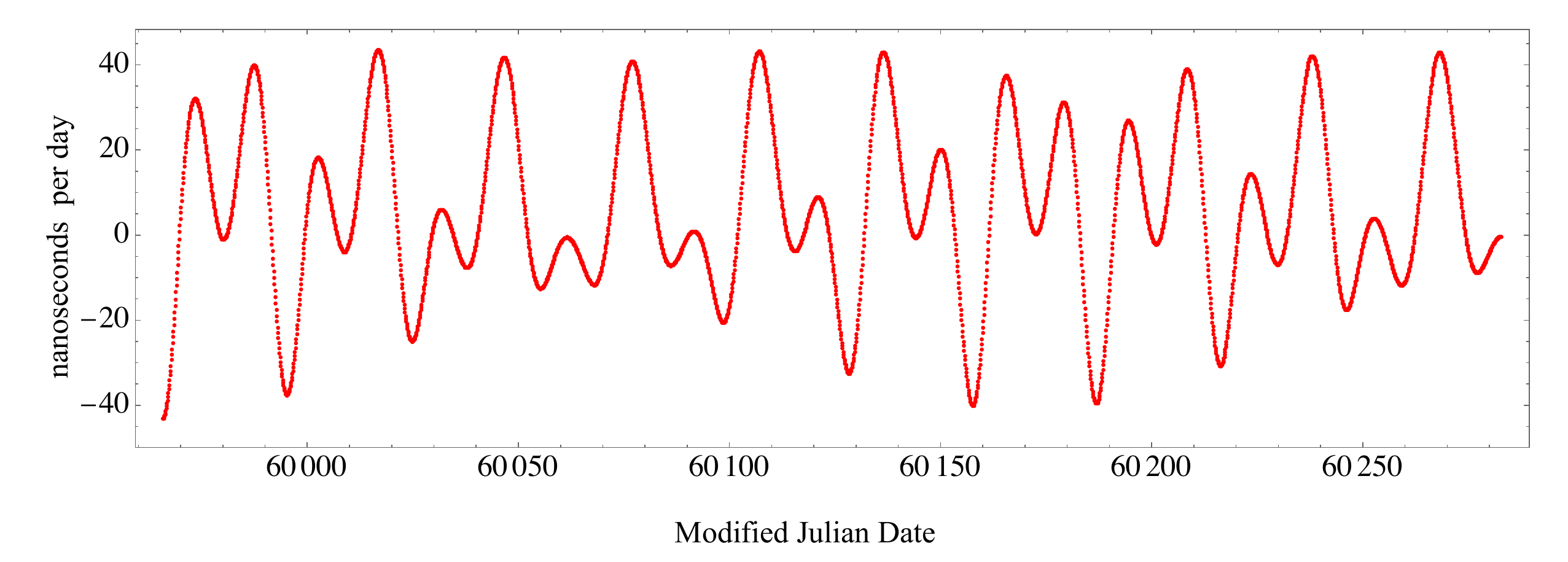}
    \caption{Time varying component of the difference between TCG and TCL. $\delta L_{Gm}$ is obtained by computing the residuals for Eq.~(\ref{def:lgm}) using DE440 and the Keplerian model. The accumulated error in rate estimation throughout a lunar orbit can be as high as $\sim 75$\,ns. When the fluctuating components are modeled and accounted for, the clocks on the Moon can be synchronized to within a few ns a day with daily steers of the order of a ns. For comparison, the timescale at the National Institute of Standards and Technology (NIST) is routinely steered, and it can be as high as 250~ps/day.  As a result, the Coordinated  Universal Time (UTC) realized locally at Boulder CO, UTC(NIST), stays within $\pm 2$~ns with respect to UTC over a year.}
    \label{fig:lgm_offset}
\end{figure}

In the inertial frame centered on Earth’s center, the Earth’s velocity is zero, and the time is defined by a standard clock at the origin---Geocentric Coordinate Time (TCG)~\cite{iau}.  A further scale change $t_{TCG}\rightarrow(1+L_G)t$, see for example Eq.~(\ref{scale_change}), defines a coordinate time whose rate is the same as that of standard clocks at rest on Earth’s geoid~\cite{petit_2005}.  Similarly, in a freely falling inertial frame centered on the Moon’s center, the Moon’s velocity is zero, and a local time is defined by a standard clock at the Moon’s origin; this could be called TCL.  A further time change such as $t_{TCL}\rightarrow(1+L_m)t$ would define a coordinate time whose rate equals that of standard clocks at rest on the Moon’s selenoid.  
Then, using Eq.~(\ref{rate_Moon}),
\ba
dt_{TCL} \approx  (1+ L_{Gm})dt_{TCG}, \quad {\rm where}\nonumber \\
L_{Gm}=\frac{(GM_m-GM_e)}{c^2 D}-\frac{1}{2 c^2}(V_m^2-V_e^2).
\label{def:lgm}
\ea
$L_{Gm}$ in Eq~(\ref{def:lgm}) is a rate with periodic contribution arising from Earth-Moon orbital eccentricity, which varies as the true anomaly ~ $-1.49373 - 0.10967 \cos(f) ~ \mu$s/day. The computed offset of $L_{Gm}$ compared with data from DE440 is given in Fig.~\ref{fig:lgm_offset}.

It might appear that the second-order Doppler contribution to the rate difference depends on the coordinate system used. In the center-of-mass coordinates, a difference of squares of velocities appears, but in a system in which the Earth is at rest, the square of the relative velocity appears. Appendix 4 shows that contributions from the centrifugal potential, which occurs in a rotating coordinate system resolve the apparent discrepancy.

\section{4.~Clocks at Earth-Moon Lagrange points}

The Lagrange points offer a cost-effective and low-noise environment for stationing spacecraft with clocks because there is no net acceleration. As a result, the orbits of clocks at these points with respect to the Earth-Moon barycenter are well-determined and the corresponding frequency offsets are precisely determined. Therefore, the proper times of clocks at Lagrange points can be related to the proper times of clocks on the Earth and  Moon with high precision. This information is crucial for synchronizing remote clocks in cislunar space. Both $L_4$ and $L_5$ are stable points, but $L_1$ and $L_2$ are only metastable locations.
\subsection{4.1~ Clock at Lagrange point $L_{1}$}

In this section, we consider a clock at $L_1$ and compare its rate to a clock on Earth's surface. It is most convenient to use a freely falling inertial frame centered at the center of mass of the Earth-Moon system for this calculation.  First, we need the precise position of $L_1$; most treatments assume circular motion, but this is inadequate for our purposes.  We assume Keplerian elliptical orbits for the Earth and Moon about the center of mass, for which the Earth-Moon distance is given by Eq.~(\ref{DtoMoon}).  The radial velocity is then
\be 
\dot D=\frac{n a e \sin(f)}{\sqrt{1-e^2}}\,,
\ee	
and the radial acceleration of the Moon relative to Earth is
\be 
\ddot  D =  \frac{GM_T e \cos(f)(1+e \cos(f))^2}{a^2(1-e^2)^2}\,.
\ee
Assume that $L_1$ is between the Earth and Moon and at a distance $x_1 D$ from the Moon, the net gravitational force towards the Moon supplies the radial acceleration of $L_1$, diminished by the centripetal acceleration due to rotation of the Earth-Moon line.  This gives the condition
\be 
-\frac{GM_e}{(D(1-x_1))^2}+\frac{GM_m}{(x_1 D)^2}=\ddot D(1-\mu-x_1)-D\left( \frac{df}{dt}\right)^2(1-\mu-x_1)\,.
\label{L1eq}
\ee
%
Solving Eq.~(\ref{L1eq}) for $x_1$  gives
\be 
 x_1 = 0.15093428(1)\,.
\ee
The metric, neglecting solar tides, is
\ba 
-ds^2=-\left(1+\frac{2\Phi_e}{c^2}+\frac{2\Phi_m}{c^2}\right)(dX^0)^2
+\left(1-\frac{2\Phi_e}{c^2}-\frac{2\Phi_m}{c^2} \right)(dX^2+dY^2+dZ^2)\,.
\ea
The motion is so slow that the potential terms in the last line can be neglected.  The transverse and radial velocities of the clock due to the rotation of the Earth-Moon line are
\ba
V_{t(L1)}^2=\left(\frac{df}{dT} \right)^2\left(\frac{M_e}{M_T}-x_1 \right)^2 D^2 ; \quad\quad 
V_{r(L1)}^2={\dot D}^2 \left(\frac{M_e}{M_T}-x_1\right)^2\,.
\ea
So, the total velocity squared is
\be\label{VL1sqrd}
V_{(L1)}^2=\frac{GM_T(1-\mu -x_1)^2}{a(1-e^2)}(1+2 e \cos(f)+e^2)\,.
\ee
The proper time elapsed during time interval $dX^0$ is
\ba 
d \tau_{L1}=\frac{dX^0}{c}\left(1-\frac{GM_e}{c^2 D(1-x_1)}-\frac{GM_m}{c^2 D x_1}
-\frac{GM_T (1-\mu -x_1)^2}{2 c^2 a(1-e^2)}(1+2 e \cos(f)+ e^2)\right)\,.
\ea
For a clock on Earth's surface,
\ba 
-ds^2=-\left(1-\frac{2 GM_e}{c^2 r_e}\left(1-\frac{J_2 a_e^2}{r_{e}^2}P_2(\cos \theta)\right)
-\frac{2 GM_m}{c^2 D}\right)(c dT)^2
+\frac{V^2}{c^2} (c dT)^2\,,\hbox to .3 truein{}
\ea
where $a_e$ is the equatorial radius of the Earth,  $J_2$ is the constant describing the oblateness of the Earth, $P_2$ is the Legendre polynomial of degree 2, and $\theta$ is the latitude on the Earth's surface at a distance $r_e$ from the center of the Earth. We approximated the distance from the Moon's center by $D$ since the position of the comparison clock on Earth's surface is unspecified.  The total velocity of the clock on Earth's surface is composed of orbital velocity plus Earth's rotational velocity:
\be 
\Vvec=\Vvec_{orbit}+\omega_e \times \bf{r_e},
\ee
\be
V^2 =V_{orbit}^2+2\Vvec_{orbit} \cdot (\omega_e \times \bf{r_e})+(\omega_e \times \bf{r_e})^2\,,
\ee
where $\omega_e$ is the rate of Earth's rotation.
We set aside the cross-term since the position of the Earth-based clock is changing rapidly, and this term averages down.  The centripetal term is grouped with the Earth potential term.  The square of the orbital velocity is composed of the squared radial velocity and the squared transverse velocity:
\ba 
V_{orbit}^2={\dot D}^2 \left(\frac{M_m}{M_T}\right)^2+\left(\frac{df}{dT} \right)^2 \left(\frac{M_m}{M_T} \right)^2 D^2\nonumber\\
=\mu^2 \frac{G M_T}{a(1-e^2)}(1+2 e \cos(f)+e^2)\,.
\ea
The Earth's potential contribution plus the rotational term can be replaced by $2\Phi_0/c^2$.  The proper time interval for the clock on Earth is then approximately
\be 
d \tau_e = \frac{dX^0}{c} \left(1+\frac{\Phi_0}{c^2}-\frac{GM_m}{c^2 D} - \mu^2 \frac{G M_T}{2 c^2a(1-e^2)}(1+2 e \cos(f)+e^2) \right),
\ee
where $cdT=dX^0$.
The fractional rate difference is then
\ba 
\frac{d\tau_{L1}-d\tau_e}{d \tau_e} =-\frac{GM_e}{c^2 D(1-x_1)}-\frac{GM_m}{c^2 D x_1}-\frac{\Phi_0}{c^2}
-\frac{G M_T (1-\mu -x_1)^2}{2c^2a(1-e^2)}(1+2 e \cos(f)+e^2)\hbox to .3 truein{}\nonumber\\
 + \frac{GM_m}{c^2 D}+\mu^2 \frac{G M_T}{2c^2a(1-e^2)}(1+2 e \cos(f)+e^2)\,.
\ea 
Evaluating this result for $x_1 = 0.150934285 $, and using values given in Table~\ref{tab:const} gives
\be 
\frac{d\tau_{L1}-d\tau_e}{d \tau_e}=6.7838449(12) \times 10^{-10}-1.2426049(12) \times 10^{-12}\cos(f)\,,
\ee
\be 
\frac{d\tau_{L1}-d\tau_e}{d \tau_e}=58.612420(12) - 0.10736106(12) \cos(f)~\mu{\rm s/day.}
\ee

The result is dominated by the term in $\Phi_0$ because the clock is high up in the Earth's potential.

\subsection{4.2.~Clock at Lagrange point $L_2$}

In this case, the Lagrange point is at a distance $x_2 D$ on the side of the Moon away from Earth. Gravitational forces due to both Earth and Moon are towards the Earth, and supply the force necessary for the centrifugal acceleration, diminished by the radial acceleration.  This yields the condition
\be
\frac{GM_e}{D^2(1+x_2)^2 }+\frac{GM_m}{x_2^2 D^2}=\frac{GM_T(1-\mu+x_2)}{D^2}\,.
\ee
which is the same as if one had assumed the orbits were circular.  Solving this equation for $x_2$, we find
\be 
x_2=0.16783274(1).
\ee
The velocity squared of the clock, composed of radial and transverse velocities squared, is
\be 
V_{(L2)}^2=\frac{GM_T(1-\mu +x_2)^2}{a(1-e^2)}(1+2 e \cos(f)+e^2)\,.
\ee
The proper time on a clock at $L_2$ is then
\ba 
d \tau_{L2}=\frac{dX^0}{c}\left(1-\frac{GM_e}{c^2 D(1+x_2)} -\frac{GM_m}{c^2 D x_2} 
-\frac{GM_T(1-\mu + x_2)^2}{2 c^2 a(1-e^2)}(1+2 e \cos(f)+e^2)\right)\,.
\ea
For a comparison clock on Earth, the analysis is the same as for the clock at $L_1$.  Therefore, the fractional rate difference is
\ba 
\frac{d\tau_{L2}-d\tau_e}{d \tau_e} =-\frac{GM_e}{c^2 D(1+x_2)}-\frac{GM_m}{c^2 D x_2}-\frac{\Phi_0}{c^2}
-\frac{G M_T (1-\mu+x_2)^2}{2c^2a(1-e^2)}(1+2 e \cos(f)+e^2)\hbox to .3 truein{}\nonumber\\
+\frac{GM_m}{c^2 D}+\mu^2 \frac{G M_T}{2c^2a(1-e^2)}(1+2 e \cos(f)+e^2) \,.
\ea 
Evaluating this result numerically gives
\be 
\frac{d\tau_{L2}-d\tau_e}{d \tau_e}=6.7846805(12) \times 10^{-10}-1.4416552(12)\times 10^{-12}\cos(f)\,,
\ee
\be 
\frac{d\tau_{L2}-d\tau_e}{d \tau_e}=58.619639(12) - 0.12445590(12) \cos(f)~\mu{\rm s/day}.
\ee
The clock beats faster since $L_2$ is farther out in Earth's gravitational potential, 
\subsection{4.3.~Clock at Lagrange point {$L_4$} or {$L_5$}}

The clock is equidistant from Earth and the Moon.  The total velocity squared is the sum of the radial velocity squared and the transverse velocity squared:
\ba
V_{(L4)}^2=(\dot D)^2+\big(\frac{df}{dT}\big)^2D^2\nonumber\\
=\frac{GM_T}{ a(1-e^2)}(1+2 e \cos(f)+e^2)\,.
\ea
The proper time on the clock is given by
\be 
-d \tau_{L4}^2=-\left(1-\frac{2 GM_e}{c^2 D}-\frac{2 GM_m}{c^2 D } -\frac{V_{(L4)}^2}{c^2}   \right) dT^2\,.
\ee
This reduces to:
\be
d \tau_{L4}=\left(1-\frac{GM_T}{ c^2 D}- \frac{GM_T}{2 c^2 a (1-e^2)}\big(1+2e\cos(f)+e^2 \big) \right)dT\,.
\ee
Analysis of the comparison clock on Earth's surface is the same as for clocks at $L_1$ or $L_2$.  The fractional rate difference is
\ba 
\frac{d\tau_{L4}-d \tau_e}{d\tau_e}=-\frac{GM_T}{ c^2 D}
-\frac{\Phi_0}{c^2}+\frac{GM_m}{c^2 D }
-\frac{G M_T}{2ac^2(1-e^2)}\left( 1-\mu^2\right)(1+2 e \cos(f)+e^2)\,.
\ea
Evaluating this result numerically gives
\be 
\frac{d\tau_{L4}-d \tau_e}{d\tau_e}=6.7948239(12) \times 10^{-10}-1.27837388(89) \times 10^{-12} \cos(f)\,,
\ee
\be 
\frac{d\tau_{L4}-d \tau_e}{d\tau_e}=58.707278(12) - 0.11045150(89) \cos(f)~\mu{\rm s/day}\,.
\ee
\section{5.~Conclusions}

We presented a model based on Keplerian orbits for establishing coordinate time on the Moon and rates of clocks at Lagrange points in cislunar space. We have used values for Keplerian orbit parameters that can be looked up; the only parameters that fit were the times of periapsis passage. The main numerical results obtained using our approach are given in Table~\ref{tab:results}. We assumed a fixed eccentricity and fixed value for the semi-major axis for the Moon's orbit around the Earth, as the present-day values for these parameters are very slowly varying~\cite{moon_fact}.
\begin{table}
    \centering
    \begin{tabular}{llc}
    \hline
    \hline
    quantity  &  location & rate ($\mu{\rm s/day}$)\\
    \hline
    $(d\tau_{m}/d\tau_e) -1$ \hspace{2.5in} & lunar surface \quad\quad\quad  &  $ 56.0199(12) - 0.10843417(89)\cos(f)$ \\    
    $(d\tau_{L_1}/d\tau_e) -1$ \hspace{1in} & $L_1$ \quad\quad\quad  &  $ 58.612420(12) - 0.10736106(12) \cos(f)$ \\
    $(d\tau_{L_2}/d\tau_e) -1$ \hspace{1in} & $L_2$ \quad\quad\quad  &  $ 58.619639(12) - 0.12445590(12) \cos(f) $\\
    $(d\tau_{L_4}/d\tau_e) -1$          &  $L_4/L_5$  & $ 58.707278(12) - 0.11045150(89) \cos(f)  $  \\
    \hline
    \end{tabular}
    \caption{Results: computed rates for various points of interest.}
    \label{tab:results}
\end{table}

The planetary ephemeris DE440 was used to calculate the potentials and velocities of Eq.~(\ref{rate_Moon_f}); the difference between the DE440 calculation and the Keplerian model calculations is only of the order of a few ns per day.  Such differences are due to tidal potentials arising from solar system bodies. Tidal effects can be readily modeled using available orbit data and added as corrections to the Keplerian model for synchronizing remote clocks on the Moon to within a few hundreds of ps or better.  Changes in time coordinate entail changes in length scale, which should be of higher order than the $c^{-2}$ effects we have considered here (see, for example, the length scale change in Eq.~(\ref{xk_app1})).

This approach is also useful in calculating time comparisons between Earth and clocks in the neighborhood of other solar system bodies such as Mars.  The available spherical harmonic gravity potential for Mars allows an estimate of the quantity $L_M$ for Mars that includes the average equatorial potential and rotational effects, analogous to $L_G$ for Earth.  In the case of Mars, the only available coordinate systems for the description of the problem are barycentric coordinates.  The Earth-Mars rate difference is dominated by the difference in the Sun's gravitational potential at the two locations.  Keplerian models, as well as computations using DE440, can be usefully compared;  this will be the subject of a future paper. Spatial transformations accompanying time transformations also remain to be examined as part of future work.

\section{Appendix 1: Fermi coordinates with origin at the center of the Moon}

The Moon's center of mass is in free fall, and therefore its path is a geodesic.  It is useful to construct Fermi coordinates with origin at this point, since then the only forces on an object in the neighborhood of the Moon due to external bodies are tidal forces.  In this coordinate system, the Christoffel symbols due to external bodies are all zero at the origin, while contributions to Christoffel symbols from the Moon itself must be ``effaced", or discarded since they are infinite and such terms cannot cause acceleration of the Moon itself. The following calculation is taken only to order $c^{-2}$.  The geodesic in question is complicated because the Earth-Moon system orbits the Sun in an approximately Keplerian orbit, while the Moon and Earth revolve around each other in a different, approximately Keplerian orbit; this latter orbit is perturbed by the Sun's tidal potential so is not known analytically.  We can still construct Fermi coordinates since many unknown quantities cancel out.  Here, we show how the metric given in Eq.~(\ref{metric_cm}) arises.
	
We work in the plane of the Sun, Earth, and Moon and denote the total gravitational potential by
\be
\Phi=\Phi_e+\Phi_s+\Phi_m\,.
\ee
the subscripts {\it e, s}, and {\it m} represent the potentials of the Earth, Sun, and Moon, respectively.  Beginning with the metric in the solar system barycentric coordinates,
\be 
-ds^2=-\left(1+\frac{2\Phi}{c^2}\right)(dX^0)^2 +\left(1-\frac{2\Phi}{c^2}\right)(dX^2+dY^2+dZ^2)\,,
\ee
where $dX^0$ is the time and $dX,dY$ and $dZ$ are the space coordinate displacements in the barycentric coordinate system. Lowercase letters will be reserved for corresponding  quantities in the local Fermi normal coordinate system.

 We give the transformation equations between barycentric coordinates and Fermi normal  coordinates with the center at the Moon as follows:\cite{ashbybertotti86}
\ba
X^0=\int_0^{x^0}\left(1-\frac{\Phi_e(m)+\Phi_s(m)}{c^2}+\frac{V(m)^2}{2 c^2}\right)dx^0
+\frac{\Vvec(m) \cdot \rvec}{c}\,;\hbox to 1.4 truein{}\\
X^k=X^k(m) +x^k\left(1+\frac{\Phi_e(m)+\Phi_s(m)}{c^2}-\frac{\Avec(m)\cdot \rvec}{c^2}\right)
+\frac{r^2 A(m)^k}{2c^2}+\frac{V(m)^k \Vvec(m) \cdot \rvec}{2 c^2}\,.
\label{xk_app1}
\ea
Here, the notation $(m)$ as in $\Vvec(m)$ represents quantities evaluated at the Moon's center of mass.  The quantity $V(m)$ is the magnitude of the Moon's velocity.  
Transformation coefficients can be derived and are:
\ba 
\parderiv{X^0}{x^0} =1-\frac{\Phi_s(m)+\Phi_e(m)}{c^2}+\frac{V(m)^2}{2 c^2}+\frac{\Avec(m) \cdot \rvec}{c^2}\,,\quad
\parderiv{X^0}{x^k} =\frac{V(m)^k}{c}\,,\quad
\parderiv{X^k}{x^0} =\frac{V(m)^k}{c}\,,\\
\parderiv{X^k}{x^j} =\delta_j^k\left(1+\frac{\Phi_s(m)+\Phi_e(m)}{c^2}-
\frac{\Avec(m)\cdot \rvec}{c^2}\right)
+\frac{V(m)^k V(m)^j}{2 c^2} -\frac{x^k A(m)^j-x^j A(m)^k}{c^2}\,.
\ea
Transformation of the metric tensor is accomplished with the usual formula:
\be 
g_{\alpha\beta}=\parderiv{X^{\mu}}{x^{\alpha}}\parderiv{X^{\nu}}{x^{\beta}}G_{\mu\nu}\,,
\label{tens_trans}
\ee
where the summation convention for repeated indices applies.  Thus, for the time-time component of the metric tensor in the freely falling frame,
\ba
g_{00}=
\left(\parderiv{X^0}{x^0}\right)^2G_{00}+\sum_{j=1}^{j=3}\left(\parderiv{X^j}{x^0}\right)^2G_{jj}\hbox to 2.5 truein{} \nonumber\\
=\left(1-\frac{\Phi_s(m)+\Phi_e(m)}{c^2}+\frac{V(m)^2}{2 c^2}+\frac{\Avec(m) \cdot \rvec}{c^2}\right)^2
(-1)\left(1+\frac{2(\Phi_e+\Phi_m+\Phi_s)}{c^2}\right)+\nonumber \\
\frac{V(m)^2}{c^2}\left(1-\frac{2(\Phi_e+\Phi_m+\Phi_s)}{c^2}\right)\,.\hbox to 1 truein{}
\ea
Expanding and keeping terms of order $c^{-2}$, 
\ba
g_{00}=-\left(1+\frac{2\Phi_m}{c^2}+\frac{2(\Phi_e+\Phi_s)}{c^2}
-\frac{2(\Phi_e(m)+\Phi_s(m))}{c^2}+\frac{2\Avec(m) \cdot \rvec }{c^2}\right)\,.
\ea
Except for the moon's potential, terms in the last line add up to the solar tidal potential, for expanding $\Phi_e+\Phi_s$ about the origin and using
\be 
A(m)^k=-\parderiv{(\Phi_e+\Phi_s)}{X^k}\,,
\ee
we find
\ba\label{ttidal}
 \frac{\Phi_e}{c^2}+\frac{\Phi_s}{c^2}-\frac{\Phi_e(m)+\Phi_s(m)}{c^2}+\frac{\Avec_{cm} \cdot \rvec}{c^2}=\frac{\Phi_t}{c^2}\nonumber\\
= -\frac{GM_e}{c^2 r_e^5}\frac{3(\rvece \cdot \rvec)^2-r_e^2 r^2}{2}
-\frac{GM_s}{c^2 R^5}\frac{3(\Rvec \cdot \rvec)^2-R^2 r^2}{2},
\label{app1_tide}
\ea
where $\Rvec$ is the vector from the Sun to the center of the Moon, $\rvece$ is the vector from the Earth to the center of the Moon, and $\rvec$ is the vector from the center of the Moon to the observation point, in local Fermi normal coordinates.  Eq. (\ref{ttidal}) gives the total tidal potential $\Phi_t/c^2$  in the vicinity of the Moon due to the Earth and the Sun. Thus
\be 
g_{00}=-\left(1+\frac{2\Phi_m}{c^2} +\frac{2\Phi_t}{c^2}\right)\,.
\ee
For the spatial component $g_{11}$ we have
\ba 
g_{11}=\left(\parderiv{X^0}{x^1}\right)^2G_{00}+\sum_{j=1}^{j=3}\left(\parderiv{X^j}{x^1}\right)^2 G_{jj} \nonumber\\
=\left(\frac{V(m)^1}{c}\right)^2(-1)+\left(\parderiv{X^1}{x^1}\right)^2 G_{11} \nonumber\\
=1-\frac{2\Phi_m}{c^2}-\frac{2\Phi_t}{c^2}\,,\hbox to .5 truein{}
\ea
where we have again expanded and kept only terms of order $c^{-2}$.  Similarly,
\be 
g_{22}=g_{33}=g_{11}\,.\hbox to .3 truein{}
\ee
The metric component $g_{12}$ is given by
\be 
g_{12}=\parderiv{X^0}{x^1}\parderiv{X^0}{x^2}G_{00}+\sum_j\parderiv{X^j}{x^1}\parderiv{X^j}{x^2}G_{jj}\,.
\ee
Keeping only terms of order $c^{-2}$, this becomes
\ba 
g_{12}=-\frac{V(m)^1 V(m)^2}{c^2}
+\bigg(\frac{V(m)^1 V(m)^2}{2 c^2}+ \frac{V(m)^2 V(m)^1}{2 c^2} \bigg)=0\,.
\ea
Similarly,
\be 
g_{13}=g_{23}=0\,.
\ee
Summarizing, the scalar invariant with origin at the Moon's center is
\ba 
-ds^2=-\left(1+\frac{2\Phi_m}{c^2}+\frac{2 \Phi_t}{c^2} \right)(dx^0)^2
+\left(1-\frac{2\Phi_m}{c^2}-\frac{2 \Phi_t}{c^2}  \right)(dx^2+dy^2+dz^2)\,.
\ea
The speed of light $c$ is a defined quantity, which does not change when transforming coordinates.  However, because the time scale changes, the length scale will also change.  A quantity such as $\Phi_m/c^2$ has been carried forward from barycentric coordinates and one might question whether it should change due to time and length scale changes.  However, such quantities are already of order $c^{-2}$ and any such changes would be of higher order and are therefore negligible.  In these coordinates, contributions to Christoffel symbols of the second kind due to
 external bodies are zero since tidal potentials have been neglected.

\section{Appendix 2:  Construction of freely falling center of mass frame}

We illustrate the method of construction of a freely-falling, locally inertial frame, by constructing such a frame at the center of mass of the Earth-Moon system, assuming this point revolves around the Sun in an elliptical Keplerin orbit.  We keep contributions only to order $c^{-2}$ and neglect tidal contributions from solar system bodies other than the Earth, Moon, and Sun. We also neglect precessions.  The Earth and Moon describe a Keplerian orbit about the center of mass in the plane determined by Earth, Moon, and Sun.  

The metric in isotropic barycentric coordinates including only the Earth, Moon, and Sun is
\ba 
-ds^2=-\left(1+\frac{2\Phi_e}{c^2}+\frac{2\Phi_m}{c^2}+\frac{2\Phi_s}{c^2}\right)(dX^0)^2
+\left(1-\frac{2\Phi_e}{c^2}-\frac{2\Phi_m}{c^2}-\frac{2\Phi_s}{c^2}\right)(dX^2+dY^2+dZ^2)
\ea
where the gravitational potentials of the Earth, Moon, and Sun are denoted by subscripts e, m, and s respectively.  We use upper case letters to denote quantities in barycentric coordinates and lower case letters for quantities in the freely falling center of the mass frame.  We are interested in a test particle at the Earth-Moon center of mass.  The local time coordinate $x^0$ is determined by the proper time on an ideal clock at the center of mass.  Consider the transformation of coordinates~\cite{ashbybertotti86}
\ba
X^0=\int_0^{x^0}\big(1-\frac{\Phi_s (cm)}{c^2}+\frac{V(cm)^2}{2 c^2}\big)dx^0
+\frac{\Vvec_{cm} \cdot \rvec}{c}\,;\\
X^k=\int_0^{x^0} \frac{ \Vvec_{cm}^k }{c} dx^0 +x^k\left(1+\frac{\Phi_s (cm)}{c^2}-\frac{\Avec_{cm}\cdot \rvec}{c^2}\right)
+\frac{r^2 A_{cm}^k}{2c^2}+\frac{V(cm)^k \Vvec_{cm} \cdot \rvec}{2 c^2}\,.
\ea
Here $\Vvec_{cm}$ and $\Avec_{cm}$ represent the velocity and acceleration of the center of mass.

The transformation coefficients are easily obtained from the above coordinate transformations and are
\ba 
\parderiv{X^0}{x^0} =1-\frac{\Phi_s(cm)}{c^2}+\frac{V(cm)^2}{2 c^2}+\frac{\Avec_{cm} \cdot \rvec}{c^2}\,,\quad
\parderiv{X^0}{x^k}=\frac{V(cm)^k}{c}\,,\quad
\parderiv{X^k}{x^0}=\frac{V(cm)^k}{c}\,,\\
\parderiv{X^k}{x^j}=\delta_j^k\left(1+\frac{\Phi_s(cm)}{c^2}-
\frac{\Avec_{cm}\cdot \rvec}{c^2}\right)
+\frac{V(cm)^k V(cm)^j}{2 c^2} -\frac{x^k A_{cm}^j-x^j A_{cm}^k}{c^2}\,.
\ea
Transformation of the metric tensor using Eq.~(\ref{tens_trans}):
the metric component $g_{00}$ in the center of mass frame, 
\ba 
g_{00}=
\left(\parderiv{X^0}{x^0}\right)^2G_{00}+\sum_{j=1}^{j=3}\left(\parderiv{X^j}{x^0}\right)^2G_{jj} \nonumber\\
=\left(1-\frac{\Phi_s(cm)}{c^2}+\frac{V(cm)^2}{2 c^2}+\frac{\Avec_{cm} \cdot \rvec}{c^2}\right)^2
(-1)\left(1+\frac{2(\Phi_e+\Phi_m+\Phi_s)}{c^2}\right) \nonumber \\
+\frac{V(cm)^2}{c^2}\left(1-\frac{2(\Phi_e+\Phi_m+\Phi_s)}{c^2}\right).
\label{earth_moon_00}
\ea
Expanding and keeping terms of order $c^{-2}$, 
\ba
g_{00}=-\left(1+\frac{2(\Phi_e+\Phi_m)}{c^2}
+ \frac{2\Phi_s}{c^2}-\frac{2\Phi_s(cm)}{c^2}+\frac{2\Avec_{cm} \cdot \rvec }{c^2}\right)\,.
\ea
The last three terms in the last line of Eq.~(\ref{earth_moon_00}) add up to twice the solar tidal potential, for expanding $\Phi_s$ about the center of mass point and using
\be 
A_{cm}^k=-\parderiv{\Phi_s}{X^k}\,.
\ee
we find
\ba 
 \frac{\Phi_s}{c^2}-\frac{\Phi_s(cm)}{c^2}+\frac{\Avec_{cm} \cdot \rvec}{c^2}
= -\frac{GM_s}{c^2 R^5}\frac{3(\Rvec \cdot \rvec)^2-R^2 r^2}{2}\,,
\ea
where $\Rvec$ is the vector from the center of the Sun to the center of mass point. We denote the solar tidal potential by 
\be 
\Phi_{st} = -\frac{GM_s}{c^2 R^5}\left(\frac{3(\Rvec \cdot \rvec)^2-R^2 r^2}{2}\right)\,.
\ee
Then
\ba 
g_{00}=-\left(1+\frac{2\Phi_e}{c^2}+\frac{2 \Phi_m}{c^2}+ \frac{2\Phi_{st}}{c^2} \right)\,.
\ea
For the spatial component $g_{11}$ we have
\ba 
g_{11}=\left(\parderiv{X^0}{x^1}\right)^2G_{00}+\sum_{j=1}^{j=3}\left(\parderiv{X^j}{x^1}\right)^2G_{jj}\hbox to 1 truein{}\nonumber\\
=\left(\frac{V(cm)^1}{c}\right)^2(-1)+\left(\parderiv{X^1}{x^1}\right)^2 G_{11}\nonumber\\
=1-\frac{2\Phi_e}{c^2}-\frac{2\Phi_m}{c^2}-\frac{2\Phi_{st}}{c^2}\,,\hbox to .5 truein{}
\ea
where we have again expanded and kept only terms of order $c^{-2}$.  Similarly,
\be 
g_{22}=g_{33}=1-\frac{2\Phi_e}{c^2}-\frac{2\Phi_m}{c^2}-\frac{2\Phi_{st}}{c^2}\,.\hbox to .3 truein{}
\ee
The metric component $g_{12}$ is given by
\be 
g_{12}=\parderiv{X^0}{x^1}\parderiv{X^0}{x^2}G_{00}+\sum_j\parderiv{X^j}{x^1}\parderiv{X^j}{x^2}G_{jj}\,.
\ee
Keeping only terms of order $c^{-2}$, this becomes
\be 
g_{12}=-\frac{V(cm)^1 V(cm)^2}{c^2}+\bigg(\frac{V(cm)^1 V(cm)^2}{2 c^2}+ \frac{V(cm)^2 V(cm)^1}{2 c^2} \bigg)=0\,.
\ee
Similarly,
\be 
g_{13}=g_
{23}=0\,.
\ee
Summarizing, the scalar invariant in the center of mass system is
\ba 
-ds^2=-\big(1+\frac{2\Phi_e}{c^2}+\frac{2 \Phi_m}{c^2}+ \frac{2\Phi_{st}}{c^2} \big)(dx^0)^2
+\big(1-\frac{2\Phi_e}{c^2}-\frac{2 \Phi_m}{c^2} - \frac{2\Phi_{st}}{c^2} \big)(dx^2+dy^2+dz^2)\,.\nonumber\\
\quad
\ea
The speed of light $c$ is a defined quantity, which does not change when transforming coordinates.  However, because the time scale changes, the length scale will also change.  A quantity such as $\Phi_e/c^2$ has been carried forward from barycentric coordinates and one might question whether it should change due to time and length scale changes.  However, such quantities are already of order $c^{-2}$ and any such changes would be of higher order and are therefore negligible.

\section{ Appendix 3: Equations of motion of Earth and Moon}
The equations of motion of the Earth and Moon should be checked to verify that, neglecting solar tidal forces, they orbit around each other in eccentric Keplerian ellipses.  The equation of motion of the Earth, using coordinate time $x^0$ as the independent variable, is
\ba
\frac{d^2x_e^i}{(dx^0)^2}+\Gamma_{\mu\nu}^i\frac{dx_e^{\mu}}{dx^0}\frac{dx_e^{\nu}}{dx^0}
-\Gamma_{\mu\nu}^0\frac{dx_e^{\mu}}{dx^0}\frac{dx_e^{\nu}}{dx^0}\frac{dx_e^i}{dx^0}=0.
\ea
The only Christoffel symbol contribution of order $c^{-2}$ is
\be 
\Gamma^i_{00}=-\frac{1}{2}\parderiv{g_{00}}{x^i}=\frac{1}{c^2}\parderiv{(\Phi_e+\Phi_m)}{x^i} \,.
\ee
This partial derivative must be evaluated at the Earth's center, which would introduce a singularity.  However, a body cannot cause the acceleration of its own center of mass so the term involving the Earth's potential must be ``effaced", or discarded.  The equation of motion of the Earth then becomes
\be 
\frac{d^2 x_e^i}{(dx^0)^2}-\frac{GM_m (x_e^i-x_m^i)}{c^2 \vert\rvec_e-\rvec_m\vert^2}=0\,.
\ee
A similar argument for the equation of the Moon gives
\be 
\frac{d^2 x_m^i}{(dx^0)^2}-\frac{GM_e (x_m^i-x_e^i)}{c^2 \vert\rvec_e-\rvec_m\vert^2}=0\,.
\ee
The center of mass of the Earth-Moon system should be at
\be 
x_{cm}^i=\frac{M_e x_e^i + M_m x_m^i}{M_T}.
\ee
Taking the corresponding linear combinates of the above equations of motion gives
\be 
\frac{d^2 x_{cm}^i}{(dx^0)^2}=0\,,
\ee
thus verifying that the center of mass of the Earth-Moon system is not accelerated in this coordinate system.

Let the vector from the center of the Earth to the center of the Moon be denoted by $\Dvec$. Then taking the difference between the above two equations of motion gives
\be 
\frac{d^2 \Dvec}{(dx^0)^2}-\frac{GM_T \Dvec}{c^2 D^3}=0\,,
\ee
where the distance between Earth and the Moon is given by Eq.~(\ref{DtoMoon}).
Then
\ba 
\dot D=\frac{n a e \sin(f)}{\sqrt{1-e^2}}\,,\hbox to 1. truein{}\\
\ddot D =\frac{GM_T e \cos(f) (1+e \cos(f))^2}{a^2(1-e^2)^2}\,.
\ea
The Earth-Moon system satisfies Kepler's equation in the plane of the Earth-Moon orbit:
\ba
\Dvec=D\{\cos(f),\sin(f)\}\,,\quad
\dot \Dvec=\frac{n a}{\sqrt{1-e^2}}\{-\sin(f),\cos(f)+e\}\,,\\
\ddot \Dvec=\frac{GM_T}{D^2}\{-\cos(f),-\sin(f) \}=\frac{GM_T \Dvec}{D^3}\,.
\ea

In summary, we have constructed a locally inertial, freely-falling frame of reference with origin at the center of mass of the Earth and Moon, and have shown that the Earth and Moon revolve about their mutual center of mass in a Keplerian orbit.  The coordinates are not normal Fermi coordinates in the sense that the Christoffel symbols of the second kind are not zero at the origin of coordinates when calculated in these coordinates.  This is because the geodesic along which the origin falls does not account for forces on a test particle at the origin due to Earth and Moon--only forces due to the Sun are accounted for.
\section{Appendix 4: Comparing results in rotating and non-rotating coordinate systems}
We calculate the fractional difference between a clock on the Moon's surface and a clock on the Earth's surface in three different coordinate systems.  These are (1) the center-of-mass locally inertial system; (2) a rotating system in which the $x-$axis is along the Earth-Moon line; and (3) a translated, rotating system in which the Earth is at the origin of coordinates and the Earth-Moon line is in the $x'$ direction.  We show that in all three coordinate systems, the fractional rate difference is the same.
A Keplerian orbit is assumed for the Earth-Moon system.  To simplify the calculations we assume that the clocks are on the surfaces of the respective bodies.  This is an approximation that can be refined when the actual positions of the clocks are specified.  


\subsection{4.1~Center-Of-Mass inertial coordinate system}
The scalar invariant in the locally inertial frame whose origin is at the center of mass of the Earth-Moon system, neglecting tidal terms, is
\ba
-ds^2 =-(1+\frac{2\Phi_e}{c^2}+\frac{2\Phi_m}{c^2})(c dT)^2 \nonumber\\
+(1-\frac{2\Phi_e}{c^2}-\frac{2\Phi_m}{c^2})(dX^2+dY^2+dZ^2)\,.
\ea
We use capital letters to denote coordinates in the center-of-mass system.  Anticipating that all velocities are small compared to the speed of light and that the calculations are carried out only to order $1/c^2$, the scalar invariant can be written
\be 
ds=\bigg(1+\frac{\Phi_e}{c^2}+\frac{\Phi_m}{c^2}-\frac{(V_x^2+V_y^2+V_z^2)}{2 c^2}\bigg) c dT\,.
\ee
For a clock on the Moon,
\ba 
V_x=\frac{d}{dt}\left(\frac{D M_e}{M_T} \right), \quad
V_y=\frac{df}{dt}\frac{D M_e}{M_T}\,.
\ea
Then using Eq. (\ref{identity}),
\be 
V_x^2+V_y^2=\frac{M_e^2 G M_T(1+2 e \cos(f)+e^2)}{2c^2M_T^2a(1-e^2)}\,.
\ee
Then the proper time on a clock on the Moon during a coordinate time interval $dT$ is
\be 
d\tau_m=\left(1-\frac{GM_e}{c^2 D}-\frac{GM_m}{c^2 R_m} - \frac{M_e^2 G M_T (1+2 e \cos(f)+e^2)}{2 c^2 M_T^2 a(1-e^2)}\right) dT\,.
\ee
For a clock on Earth, 
\ba 
V_x=\frac{d}{dt}\left(-\frac{D M_m}{M_T}\right), \quad
V_y=-\frac{df}{dt}\left(\frac{D M_m}{M_T}\right)\,.
\ea
Then 
\be 
V_x^2+V_y^2=\frac{M_m^2 G M_T(1+2 e \cos(f)+e^2)}{M_T^2 a(1-e^2)}\,.
\ee
and the proper time elapsed during a coordinate time interval $dT$ is
\be
d\tau_e=\big(1-\frac{GM_e}{c^2 R_e}-\frac{G M_m}{c^2 D} - \frac{M_m^2 G M_T (1+2 e \cos(f)+e^2)}{2 c^2 M_T^2 a(1-e^2)}\big)dT.
\ee
The fractional difference is
\ba\label{fractdiff} 
\frac{d\tau_m -d\tau_e}{d\tau_e}=-\frac{GM_e}{c^2 D}-\frac{GM_m}{c^2 R_m}+\frac{GM_e}{c^2 R_e}+\frac{GM_m}{c^2 D}
-\frac{(M_e^2-M_m^2)(1+2 e \cos(f)+e^2)}{2 c^2 M_T^2a(1-e^2)}\,.\hbox to .5 truein{}
\ea
The difference in the last term represents a difference of squares of velocities.
\subsection{4.2~Rotating center-of-mass coordinates}

	Introduce a rotating system with an Earth-Moon line along the new x-axis:
\ba 
X=x \cos(f)-y \sin(f)\,;\nonumber\\
Y=x \sin(f)+y \cos(f)\,;\nonumber\\
T=t\,.\hbox to 0.9 truein {}
\ea
Then
\be 
dX^2+dY^2=dx^2+dy^2 + \frac{df}{dt}^2(x^2+y^2)dt^2+2\frac{df}{dt} dt (xdy-ydx)\,.
\ee
The scalar invariant becomes
\ba 
-ds^2=-\left((1+\frac{2 \Phi_e}{c^2}+\frac{2 \Phi_m}{c^2}-\big(\frac{df}{dt}\big)^2 (x^2+y^2)\right)(c dt)^2\nonumber\\
+ 2\frac{df}{dt} dt (x dy-y dx) + dx^2+dy^2+dz^2\,.\hbox to .2 truein{}
\ea
For a clock on the Moon,
\ba 
x_m=\frac{M_e D}{M_T}\,,\quad y_m=0\,;\quad\quad
\dot x_m=\frac{M_e \dot D}{M_T}\,,\quad \dot y_m=0\,.
\ea
For a clock on the Moon, there is no contribution from the Sagnac term.  The proper time interval is
\ba 
d\tau_m=\left(1-\frac{GM_e}{c^2 D}-\frac{GM_m}{c^2 R_m}-\left(\frac{df}{dt}\right)^2 \frac{M_e^ 2 D^2}{2 c^2 M_T^2}
-\frac{M_e^2 \dot D^2}{2 c^2M_T^2} \right)dt\hbox to .5 truein{}\nonumber\\
=\left(1-\frac{GM_e}{c^2 D}-\frac{GM_m}{c^2 R_m}
-\frac{M_e^2 G M_T(1+2 e \cos(f)+e^2)}{2 c^2 a (1-e^2)M_T^2}\right)dt 
 \ea
Note that there is a significant contribution from the centrifugal potential.  For a clock on Earth,
\ba 
x=-D\frac{M_m}{M_T};\quad \dot x=-\dot D\frac{M_m}{M_T}; \quad\quad
y=0;\quad \dot y=0\,.
\ea
The proper time interval is then
\ba 
d\tau_e=\left(1-\frac{GM_e}{c^2 R_e}-\frac{GM_m}{c^2 D}
-\left(\frac{df}{dt} \right)^2\frac{1}{2c^2}\left(-D\frac{M_m}{M_T}\right)^2
-\frac{1}{2c^2}\left(-\dot D\frac{M_m}{M_T} \right)^2 \right)dt\,.
\ea
The fractional proper time interval difference reduces exactly to the expression in Eq. (\ref{fractdiff}).

\subsection{4.3~Rotating coordinates with Earth at origin}

For this system, the velocity of the Moon is the relative velocity.  This implies the use of a coordinate system in which the Earth is not moving.  This has to be a rotating coordinate system with its origin coinciding with the Earth's center. Therefore translating the origin to the center of the Earth, with no change in the time variable,
\ba 
x=x'-D\frac{M_m}{M_T}; \quad dx=dx'-\dot D dt \frac{M_m}{M_T}\,;\nonumber\\
y=y';\quad dy=dy';\quad z=z'\,.\hbox to 0.75 truein{}
\ea
The scalar invariant becomes
\ba 
-ds^2=-\left(1+\frac{2\Phi_e}{c^2}+\frac{2 \Phi_m}{c^2}
-\big(\frac{df}{c dt}\big)^2 \big((x'-D\frac{M_m}{M_T})^2+y'^2 \big) \right)(c dt)^2\nonumber\\
+2\frac{df}{dt} dt \left((x'-D\frac{M_m}{M_T})dy'-y'(dx'-\dot D \frac{M_m}{M_T} dt) \right)\nonumber\\
+(1+...)\left((dx'-\dot D \frac{M_m}{M_T} dt)^2 +dy'^2+dz'^2 \right)\,.
\ea
The potentials in the last term have been suppressed since they do not contribute to the order of this calculation.  For a clock on the surface of the Moon, 
\ba 
x'=D; \quad \dot x'=\dot D \quad ({\rm radial\ velocity})\nonumber\\
y'=0;\quad \dot y'=0\,.\hbox to .5 truein {}
\ea
There is no contribution from the Sagnac term but there is a significant contribution from the centrifugal potential, representing the transverse velocity of the Moon.  The radial velocity of the Moon comes from the spatial part of the metric.  The proper time interval for such a clock is
\ba
d\tau_m=\left( 1-\frac{GM_e}{c^2 D}-\frac{GM_m}{c^2 R_m}-\frac{1}{2 c^2}\left(\frac{df}{dt} \right)^2\left(D-D\frac{M_m}{M_T}\right)^2\right)dt
+\frac{1}{2c^2}\left(\dot D-\dot D \frac{M_m}{M_T}\right)^2 dt\hbox to 1 truein{}\nonumber\\
=\left(1-\frac{GM_e}{c^2 D}-\frac{GM_m}{c^2 R_m}
-\frac{M_e^2 G M_T(1+2 e \cos(f)+e^2)}{2 c^2 a (1-e^2)M_T^2}\right)dt. \hbox to 0.6 truein{}
\ea
For a clock on the surface of the Earth, 
\be 
x'=y'=\dot x'=\dot y' =0\,.
\ee
The proper time interval is 
\ba 
d\tau_e=\left(1-\frac{GM_e}{c^2 R_e}-\frac{GM_m}{c^2 D}
-\left(\frac{df}{dt} \right)^2\frac{1}{2c^2}\left(-D\frac{M_m}{M_T}\right)^2
-\frac{1}{2c^2}\left(-\dot D\frac{M_m}{M_T} \right)^2 \right)dt\nonumber\\
=\left(1-\frac{GM_e}{c^2 R_e}-\frac{GM_m}{c^2 D}
- \frac{M_m^2 G M_T (1+2 e \cos(f)+e^2)}{2 c^2 M_T^2 a(1-e^2)}\right)dt.
\ea
It is easily seen that the fractional proper time difference reduces to expressions that have been previously derived.  Thus in all three coordinate systems, the fractional proper time difference is the same.

\subsection*{Acknowledgments}
We would like to acknowledge the funding we received from the NASA grant NNH12AT81I. We are grateful to Elizabeth Donley, who carefully and critically reviewed the manuscript and provided valuable suggestions. We would also like to express our gratitude to Cheryl Gramling for initiating discussions on lunar time. We extend our sincere thanks to Roger Brown, Thomas Heavner, Judah Levine, Jeffrey Sherman, and Daniel Slichter for their review of the manuscript. This work is a contribution of NIST and is not subject to US copyright.

\end{document}